\documentclass{emulateapj}
\usepackage{apjfonts}
\usepackage{epsf}
\newcommand{\Msun}{\mathrm{M}_{\odot}}
\newcommand{\kms}{\mathrm{km\ s}^{-1}}
\newcommand{\gamtr}{\gamma_{\rm tr}}
\newcommand{\gamsig}{\gamma_{\sigma}}

\begin{document}

\slugcomment{Accepted to ApJ Letters}
\shortauthors{Gnedin et al.}
\shorttitle{Mass Profile of the Galaxy}

\title{The Mass Profile of the Galaxy to 80 kpc}

\author{ Oleg Y. Gnedin\altaffilmark{1},
         Warren R. Brown\altaffilmark{2},
         Margaret J. Geller\altaffilmark{2},
         Scott J. Kenyon\altaffilmark{2}}

\altaffiltext{1}{Department of Astronomy, University of Michigan, 
   Ann Arbor, MI 48109; \mbox{\tt ognedin@umich.edu}}
\altaffiltext{2}{Smithsonian Astrophysical Observatory, Cambridge, MA 02138;
   \mbox{\tt \{wbrown,mgeller,skenyon\}@cfa.harvard.edu}}

\date{\today}

\begin{abstract}
The Hypervelocity Star survey presents the currently largest sample of
radial velocity measurements of halo stars out to 80 kpc.  We apply
spherical Jeans modeling to these data in order to derive the mass
profile of the Galaxy.  We restrict the analysis to distances larger
than 25 kpc from the Galactic center, where the density profile of
halo stars is well approximated by a single power law with logarithmic
slope between $-3.5$ and $-4.5$.  With this restriction, we also avoid
the complication of modeling a flattened Galactic disk.  In the range
$25 < r < 80$ kpc, the radial velocity dispersion declines remarkably
little; a robust measure of its logarithmic slope is between $-0.05$
and $-0.1$.  The circular velocity profile also declines remarkably
little with radius.  The allowed range of $V_c\, (80\, \mathrm{kpc})$
lies between 175 and 231 $\kms$, with the most likely value 193
$\kms$.  Compared with the value at the solar location, the Galactic
circular velocity declines by less than 20\% over an order of
magnitude in radius.  Such a flat profile requires a massive and
extended dark matter halo.  The mass enclosed within 80 kpc is
$6.9^{+3.0}_{-1.2} \times 10^{11}\ \Msun$.  Our sample of radial
velocities is large enough that the biggest uncertainty in the mass is
not statistical but systematic, dominated by the density slope and
anisotropy of the tracer population.  Further progress requires
modeling observed datasets within realistic simulations of galaxy
formation.
\end{abstract}


\section{Introduction}
 \label{sec:intro}

Measuring the mass of the Galaxy is an astonishingly difficult task.
Convenient tracers of disk rotation -- stars and gas clouds -- extend
only to 20 kpc \citep[e.g.][]{sofue_etal09}.  At larger radii,
statistical analysis of radial velocities must be used.  Traditional
tracers at distances up to 100 kpc include globular clusters and
dwarf satellite galaxies \citep[e.g.,][]{kochanek96,
  wilkinson_evans99, sakamoto_etal03}.  Recently,
\citet{battaglia_etal05} derived the radial velocity dispersion
profile to 120 kpc using a combined sample of globular clusters,
satellite galaxies, and halo red giant stars.  They found the
dispersion falling from $\sim 120\ \kms$ to $\sim 50\ \kms$ between 20
and 120 kpc.  In contrast, \citet{xue_etal08} assembled a large sample
of blue horizontal-branch (BHB) stars from the Sloan Digital Sky
Survey Data Release 6 (SDSS DR6) and found a much flatter profile
between 20 and 60 kpc.  Here we use a new spectroscopic survey
\citep{brown_etal10a} aimed at finding hypervelocity stars (HVS) to set
the most precise constraint on the Galactic mass within 80 kpc.

\citet{brown_etal10a} present a sample of 910 late B-stars and early
A-stars in the Galactic halo.  Their luminosity, and therefore
distance, depends on whether these stars are BHB stars or
main-sequence blue stragglers with similar effective temperature and
surface gravity.  The ambiguous nature of the stars is especially
problematic at redder colors, $u-g > 0.6$, where the luminosity can
differ by a factor of 5, as demonstrated in Fig. 4 in
\citet{brown_etal10a}.  This bimodal distribution of distance does not
have a well-defined average value and therefore requires statistical
sampling.  \citet{brown_etal10a} create 100 Monte Carlo realizations of
each star to sample the color and metallicity distributions and to
derive the distributions of luminosity and distance.  We use this full
Monte Carlo catalog of distances in our analysis, while retaining the
observed values of radial velocity and its uncertainty.

\section{Method}

We use the spherical Jeans equation \citep{binney_tremaine08} to solve
for the circular velocity $V_c$ given the radial velocity dispersion
$\sigma_r$ of tracer particles:
\begin{equation}
  V_c^2 = {G M(r) \over r} = \sigma_r^2 \left({ 
    -{d\ln{\rho_{\rm tr}} \over d\ln{r}} -{d\ln{\sigma_r^2} \over d\ln{r}} -2\beta }
    \right),
  \label{eq:jeans}
\end{equation}
where $M(r)$ is the enclosed total mass within radius $r$, $\rho_{\rm
tr}$ is the density of tracer population, and $\beta$ is the
anisotropy parameter.  This equation assumes that the mass
distribution $M(r)$ is static and spherically symmetric.

In our case the tracers are halo stars, which follow a steep density
profile with negative logarithmic slope (see below)
\begin{equation}
  \gamtr \equiv -{d\ln{\rho_{\rm tr}} \over d\ln{r}} \approx 4.
\end{equation}
The velocity dispersion slope and the anisotropy parameter are
typically less than unity, and therefore, $\gamtr$ dominates the last
factor in equation (\ref{eq:jeans}).  The circular velocity profile
depends mainly on $\sigma_r$ and $\gamtr$.

For ease of comparison with mass profiles expected from cosmological
simulations, we fit a power law relation to the radial velocity
dispersion, normalized at radius $r_0$:
\begin{equation}
  \sigma_r(r) = \sigma_0 \left({ r / r_0 }\right)^{-\gamsig}.
  \label{eq:sigmar}
\end{equation}
We apply a maximum likelihood (ML) method, similar to that described
in Appendix of \citet{gnedin_etal07}, to maximize the probability of
the model fit given the observations:
\begin{equation}
  {\cal L} = \prod_i {1 \over \sqrt{2\pi (\sigma_r^2(r_i) + \sigma_{v,i}^2)}}
   \exp{\left[{-{1\over 2} {v_{r,i}^2 \over \sigma_r^2(r_i) + \sigma_{v,i}^2}}\right]}.
\end{equation}
We assume that each radial velocity $v_{r,i}$ is drawn from a Gaussian
distribution with zero mean and with combined variance of the
intrinsic dispersion at its location, $\sigma_r(r_i)$, and the
measurement uncertainty $\sigma_{v,i}$.  The ML method uses all
available observational information without binning the data.

The value of the velocity dispersion is sensitive to the presence of
outliers in the sample.  In order to remove them, we impose a cut
$|v_{r,i}| \le V_{\rm esc}$, with the escape velocity calculated
self-consistently from the derived mass profile.  In the
``most-likely'' case this process removes only 6 out of 558 stars
located at $r > 25$ kpc.

\begin{figure}[t]
\vspace{-0.3cm}
\centerline{\epsfxsize3.5truein \epsffile{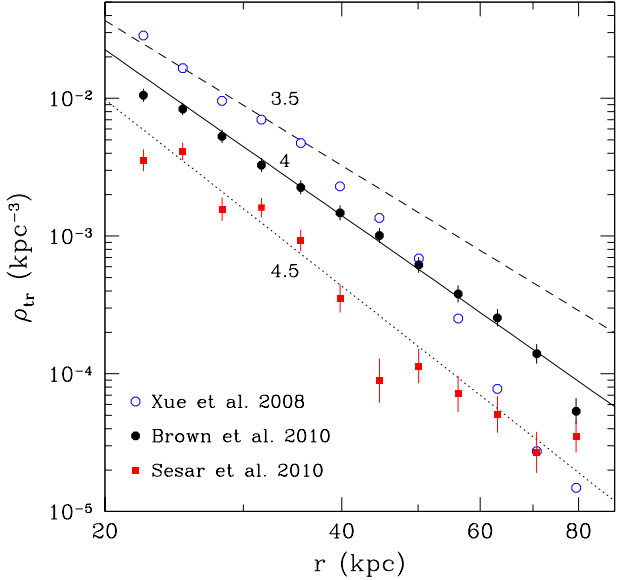}}
\vspace{-0.3cm}
\caption{Density profile of tracer populations: BHB stars from HVS
  survey ({\it solid circles}), BHB stars from SDSS DR6 ({\it open
  circles}), and RR Lyrae from SDSS Stripe 82 ({\it solid squares}).
  The density is calculated directly from observed counts, without
  correcting for sky incompleteness.  The true density is a factor
  $\sim 5$ higher.  Errors are from Poisson statistics.
  Lines show best fits for the logarithmic slope, from 3.5 ({\it
  dashed}) to 4 ({\it solid}) to 4.5 ({\it dotted}).}
\vspace{0.3cm}
  \label{fig:den}
\end{figure}

The escape velocity at a given radius $r$ measures the total
gravitational potential, a sum of the two terms interior and
exterior to $r$:
\begin{equation}
  V_{\rm esc}^2 \equiv -2 \Phi(r) 
    = {2 G M(r) \over r} + 8\pi G \int_r^\infty dr' r' \rho(r').
\end{equation}
We evaluate the second term by adopting a power-law profile for the
total matter density $\rho(r) \propto r^{-2-\alpha}$, out to a maximum
radius $r_{\rm out}$.  In the outer Galactic halo we expect $0 <
\alpha < 1$.  Then the escape velocity is
\begin{equation}
  V_{\rm esc}^2 = 2 V_c^2 + {2 G \over \alpha} 
    \left({{dM \over dr} - {{dM \over dr}\mid_{\rm out}}}\right).
\end{equation}
Ignoring the boundary condition at $r_{\rm out}$, we obtain 
\begin{equation}
  V_{\rm esc}^2 \le {2 \over \alpha}\; V_c^2.
\end{equation}
\citet{watkins_etal10} show that most halo tracers reside in the
radial range where $\alpha \approx 0.5$, and thus $V_{\rm esc}(r)
\approx 2 \, V_c(r)$.  As the resulting circular velocity profile is
close to flat, we adopt this simple relation in our analysis.  The ML
fit converges to the same value regardless of the initial guess for
$V_{\rm esc}$.  In the inner regions of the Galaxy, lower $\alpha$ 
leads to higher escape velocity, consistent with the estimate
from the RAVE survey of $V_{\rm esc} \approx 500-600\ \kms$ at the
solar location \citep{smith_etal07}.

The density profile of RR Lyrae halo stars shows a break around
$25-30$ kpc \citep{watkins_etal09, sesar_etal10}, with shallow inner
slope ($\gamtr \approx 3$) and steep outer slope ($\gamtr \approx
4.5$).  The stellar halo is also less flattened in the outer regions.
In order to obtain a robust measure of the mass profile, we restrict
our analysis only to radii $r > 25$ kpc.  Figure~\ref{fig:den} shows
that at these radii the density profiles of both BHB stars and RR
Lyrae are consistent with our fiducial value $\gamtr \approx 4$.

We consider the following parameter ranges: $3.5 \le \gamtr \le 4.5$
and $0 \le \beta \le 0.5$.  For each parameter set, we use the full
Monte Carlo catalog and self-consistently remove outliers using the
value of the escape velocity resulting from the velocity dispersion
profile.  This procedure gives simultaneous estimates for both
$\sigma_r(r)$ and $V_c(r)$.

\begin{table}[t]
\begin{center}
\caption{\sc Characteristic Parameters of the Mass Profile}
\label{tab:obs}
\begin{tabular}{lllccc}
\tableline\tableline\\
\multicolumn{1}{l}{Scenario} &
\multicolumn{1}{c}{$\gamtr$} &
\multicolumn{1}{c}{$\beta$} &
\multicolumn{1}{c}{$\sigma_0$ ($\kms$)} &
\multicolumn{1}{c}{$\gamma_\sigma$} &
\multicolumn{1}{c}{$V_c(80)$ ($\kms$)}
\\[2mm] \tableline\\
Smallest mass    & 3.5 & 0.5 & 109 & 0.077 & 168 \\
More likely, min & 3.5 & 0.4 & 110 & 0.089 & 175 \\
{\bf Most likely} & {\bf 4} & {\bf 0.4} & {\bf 111} & {\bf 0.078} & {\bf 193} \\
More likely, max & 4   & 0   & 118 & 0.049 & 231 \\
Largest mass     & 4.5 & 0   & 121 & 0.088 & 246 \\
\tableline
\end{tabular}
\end{center}
\vspace{0.3cm}
\end{table}

\section{Results}

The radial velocity dispersion profile declines remarkably little with
radius.  For the power-law fit (eq. [\ref{eq:sigmar}]) we obtain
$\sigma_0 \approx 114\ \kms$ and $\gamsig \approx 0.1$ for $r_0 = 40$
kpc.  This choice of $r_0$ minimizes the error of $\sigma_0$.  We do
not quote formal errors of the fit because the uncertainty is
dominated by the unknown parameters $\gamtr$ and $\beta$, as we
discuss below.  Table~\ref{tab:obs} summarizes the parameters of the
best-fit and the most extreme allowed models.

\citet{brown_etal10a} consider a linear fit to the same data and
obtain $\sigma_r \approx 120 - 0.3\, r$, where $r$ is in kpc and
$\sigma_r$ is in $\kms$.  By linearly expanding
equation~(\ref{eq:sigmar}) near $r_0$, we derive a similar expression
in the range $25 < r < 80$ kpc: $\sigma_r \approx 120 - 0.22\, r$.

\begin{figure}[t]
\vspace{-0.3cm}
\centerline{\epsfxsize3.5truein \epsffile{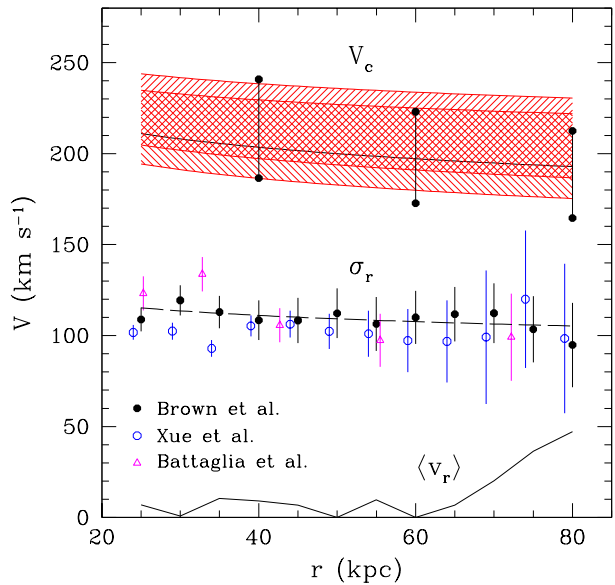}}
\vspace{-0.3cm}
\caption{Radial velocity dispersion profile of BHB stars from HVS
  survey ({\it solid circles}), from SDSS DR6 ({\it open circles},
  offset by 1 kpc for clarity), and from a combined sample of globular
  clusters, satellite galaxies, and halo red giants ({\it triangles}).
  Lond-dashed line is a power-law fit, $\sigma(r) = 111 \,
  (r/40)^{-0.08}\ \kms$.  Solid line shows the mean streaming radial
  velocity for the HVS sample.  Shaded regions illustrate the
  allowed range of circular velocity when the anisotropy parameter is
  varied from $\beta=0$ to $\beta=0.5$ for a fixed tracer density
  $\gamtr=4$ ({\it bottom left towards top right}) and when the tracer
  density is varied from $\gamtr=3.5$ to $\gamtr=4.5$ for a fixed
  $\beta=0.4$ ({\it top left towards bottom right}).  Middle solid
  line is for $\gamtr=4$, $\beta=0.4$.  Filled circles connected by
  vertical lines show the \citet{watkins_etal10} mass estimator
  applied to the HVS sample at 40, 60, and 80 kpc.}
\vspace{0.3cm}
  \label{fig:disp}
\end{figure}

\citet{xue_etal08} fit an exponential function to their SDSS DR6
sample and obtain $\sigma_r = 111\, \exp(-r/354\, \mathrm {kpc})\
\kms$.  A linear approximation gives $\sigma_r \approx 111 - 0.31\,
r$, which has a similar slope and slightly lower normalization than
the HVS sample.  Since our dispersion profile is calculated from the
Monte Carlo catalog, we similarly bootstrap the \citet{xue_etal08}
sample 30 times and obtain a distance distribution, from which we
calculate the dispersion plotted on Fig.~\ref{fig:disp}.

The profiles from both samples are less steep than that derived by
\citet{battaglia_etal05}: $\sigma_r \approx 132 - 0.6\, r$.

\begin{figure}[t]
\vspace{-0.3cm}
\centerline{\epsfxsize3.5truein \epsffile{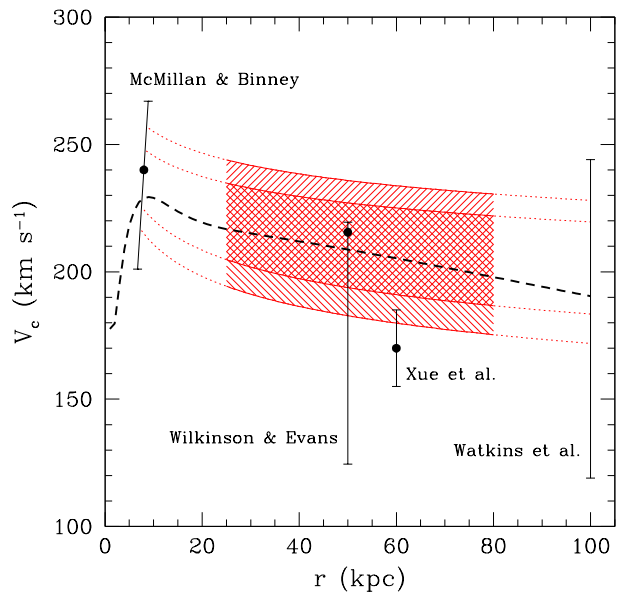}}
\vspace{-0.3cm}
\caption{ Shaded regions illustrate the allowed range of circular
  velocity when the anisotropy parameter is varied from $\beta=0$ to
  $\beta=0.5$ for a fixed tracer density $\gamtr=4$ ({\it bottom left
  towards top right}) and when the tracer density is varied from
  $\gamtr=3.5$ to $\gamtr=4.5$ for a fixed $\beta=0.4$ ({\it top left
  towards bottom right}), same as in Fig.~\ref{fig:disp}.  Thick
  dashed line shows a representative fit for an NFW halo model (see
  text for details).  Independent constraints at 8, 50, 60, and 100
  kpc are from \citet{mcmillan_binney10}, \citet{wilkinson_evans99},
  \citet{xue_etal08}, and \citet{watkins_etal10}, respectively.}
\vspace{0.3cm}
  \label{fig:vc}
\end{figure}

Figure~\ref{fig:disp} shows the velocity dispersion profiles for all
three samples.  They appear consistent with each other within the
errors, and the differences in the derived best fit parameters can be
attributed to small sample sizes.  Outside 25 kpc, the Battaglia, Xue,
and Brown samples contain 80, 741, and 558 objects, respectively
(using the distance distribution from the Monte Carlo catalogs).
Outside 50 kpc, the three samples contain 24, 76, and 163 objects,
respectively.  Note that our sample more than doubles the number of
distant stars and thus presents the most accurate current measurement
of $\sigma_r$.  Individual velocity errors and resampling of the
distance distribution affect the value of $\sigma_r$ by less than 3\%.
The streaming velocity $\langle v_r \rangle$ is small everywhere
except at the outermost radii.  Overall, the data indicate that the
velocity dispersion varies little with radius out to 80 kpc.

We can now combine equations~(\ref{eq:jeans}) and (\ref{eq:sigmar}) to
calculate the circular velocity
\begin{equation}
  V_c(r) = \sigma_0 \left({r/r_0}\right)^{-\gamsig} (\gamtr + 2\gamsig -2\beta)^{1/2}.
  \label{eq:vc}
\end{equation}
This estimate is degenerate with respect to the density and anisotropy
of the tracer population.  Lower $\gamtr$ and higher $\beta$ result in
the lower mass estimate, and vice versa.  At the same time, steeper
density slope can balance stronger anisotropy.  Based on our
derivation of the tracer density in Figure~\ref{fig:den}, we believe
the slope is constrained to be between 3.5 and 4.5, with the most
likely value of $\gamtr \approx 4$.  The anisotropy parameter is not directly
known, but we can take the predictions of cosmological simulations of
galaxy formation as a guide.  The centers of halos are close to being
isotropic.  In the outer parts, the orbits of dark matter particles
and satellite halos become more radially biased with distance,
reaching $\beta \approx 0.5$ outside the peak of the circular velocity
curve \citep[e.g.,][]{diemand_etal07b, navarro_etal10}.  Thus a full
range of values $0 \le \beta \le 0.5$ is possible, while the most
likely value in our radial range is $\beta \approx 0.4$.

Figure~\ref{fig:disp} illustrates the allowed range of circular
velocity.  We take the best fit ($\gamtr=4$, $\beta=0.4$) and consider
the full range of variation of either parameter while keeping the
other fixed.  The overlap of the two shaded regions gives the most
probable values of $V_c$.  The uncertainty due to either parameter is
systematic in nature and comparable in magnitude.  The value of the
circular velocity at 80 kpc is uncertain at least by 10\%, and the
value of the enclosed mass at least by 20\%.  The most likely values
are $V_c\, (80) = 193\ \kms$, $M(80) = 6.9\times 10^{11}\ \Msun$.

For comparison, we also use an independent robust mass estimator
proposed by \citet{watkins_etal10}.  The mass within radius $r$ is
given by the following average of radial velocities of all objects
inside $r$:
\begin{equation}
  V_c^2(r) = (\gamtr + \alpha -2\beta) \;
    \langle{ v_{r,i}^2 \left({r_i \over r}\right)^\alpha }\rangle,
\end{equation}
where $(-2-\alpha)$ is again the logarithmic slope of the density
profile in the range of radii probed by the data.  This estimator also
depends on $\gamtr$ and $\beta$, similarly to equation (\ref{eq:vc}).
In fact, if these two parameters and $\alpha$ are constant with radius,
then $\alpha = 2\gamsig$.  The estimator is different in that it takes
a single average of all radial velocities instead of fitting the
dispersion profile.

Three vertical lines in Figure~\ref{fig:disp} show the applications of
the robust estimator at 40, 60, and 80 kpc.  They indicate a slightly
steeper decline of $V_c$ with radius, but still fully consistent
with the results derived from the dispersion.  The most likely value at
80 kpc is $V_{c,est}\, (80) \approx 190\ \kms$.

Figure~\ref{fig:vc} compares our derived circular velocity with other
independent observational constraints.  Dotted lines show the
extrapolation of $V_c$ outside the range of our data and should be
treated with caution.  \citet{watkins_etal10} estimate the mass of the
Galaxy at $r=100$ kpc using the sample of all known satellite
galaxies.  Depending on the anisotropy and the inclusion of particular
galaxies, the mass can vary between 3.3 and $13.8\times 10^{11}\ \Msun$,
corresponding to the circular velocity range from 119 to 244 $\kms$.
Our results favor the upper half of this range.

Within 60 kpc, \citet{xue_etal08} estimate $V_c = 170 \pm 15\ \kms$ by
matching observed velocity dispersion to the motion of particles in
two hydrodynamic simulations of galaxy formation.  Within 50 kpc,
\citet{wilkinson_evans99} obtain $M(50) = 5.4^{+0.2}_{-3.6} \times
10^{11}\ \Msun$ by modeling the distribution function of observed
velocities.  For clarity, we do not show similar results obtained by
\citet{kochanek96}, $M(50) = 4.9^{+1.1}_{-1.1} \times 10^{11}\ \Msun$,
and \citet{sakamoto_etal03}, $M(50) = 5.5^{+0.1}_{-0.4} \times
10^{11}\ \Msun$.

The circular velocity at the solar circle, $V_0$, is constrained
better but it scales with the distance to the Galactic center, $R_0$.
\citet{mcmillan_binney10} analyze the motion of masers in star-forming
regions throughout the Galaxy and find that the best-constrained
parameter is the ratio $V_0/R_0 = 30 \pm 2\ \kms\ \mathrm{kpc}^{-1}$.
The distance $R_0$ can vary from 6.7 to 8.9 kpc, depending on model
assumptions.  We plot this constraint on $V_0$ as a diagonal line in
Figure~\ref{fig:vc}.  This line matches the extrapolation of our fits
perfectly.  However, this result should not be considered as an
improvement on the value of $V_0$, because we do not expect $V_c(r)$
to remain a power law with fixed slope at such small radii.

For illustration, we also show an example of a three-component model
of the Galactic potential, similar to the cosmologically-motivated model
by \citet{klypin_etal02}.  We assume an axisymmetric exponential disk
with the mass $5\times 10^{10}\ \Msun$ and scale length 3 kpc, and a
compact bulge with the mass $5\times 10^9\ \Msun$
\citep{binney_tremaine08}.  We include a dark matter halo represented
by an NFW profile with the scale radius $r_s = 20$ kpc and the mass
that, along with the disk and the bulge, gives the total virial mass
$M_{\rm vir} = 1.6\times 10^{12}\ \Msun$.  This model lies in the
middle of our inferred interval of circular velocity and can be taken
as a good first approximation to the Galactic mass distribution.  Note
however that the uncertainty in the value of the virial mass is at
least 20\% and possibly larger.

\begin{figure}[t]
\vspace{0cm}
\centerline{\epsfxsize3.5truein \epsffile{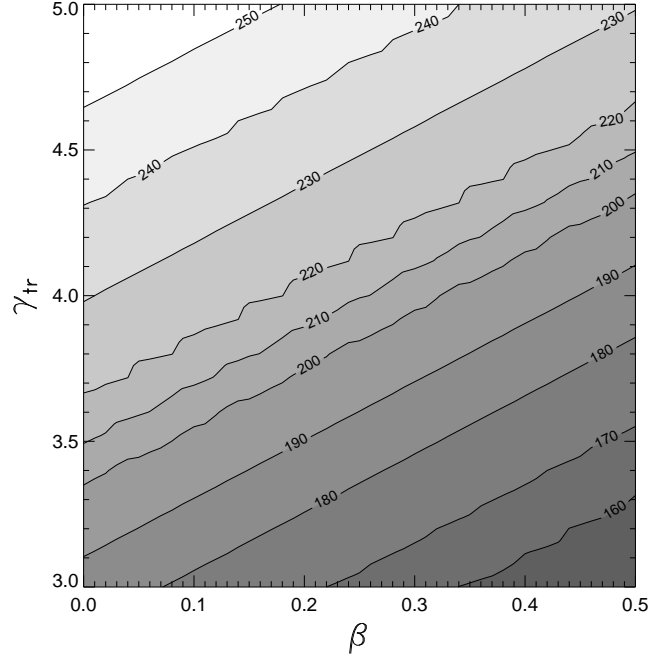}}
\vspace{0.2cm}
\caption{Contours of the circular velocity at 80 kpc as a function of
  the anisotropy parameter $\beta$ and logarithmic slope of the
  density of tracers $\gamtr$.  The values of $V_c\, (80)$ at each point
  are calculated for the ML fit to the velocity dispersion profile.}
\vspace{0.3cm}
  \label{fig:vc80}
\end{figure}

In Figure~\ref{fig:vc80} we illustrate further the dependence of the
derived value of the circular velocity at 80 kpc on the parameters
$\gamtr$ and $\beta$.  For each combination of the parameters, we run
a ML fit to the velocity dispersion profile with self-consistent
removal of unbound stars.  The plot shows, however, that contours of
$V_c\, (80)$ lie close to the lines $\gamtr - 2\beta =
\mathrm{const}$, as expected from equation (\ref{eq:vc}) for a
constant $\gamsig$.  Larger $\beta$ gives smaller velocity, but
according to cosmological simulations it should not exceed 0.5 and be
near 0.4.  However, the estimate of $V_c$ would be higher if the
tracers had tangential anisotropy, $\beta < 0$.  A density slope
$\gamtr \approx 4$ is most likely, as it fits both the BHB and RR
Lyrae samples.

\section{Summary}

Using maximum-likelihood analysis of a new sample of radial velocities
of distant halo stars, we infer that their radial velocity dispersion
profile declines little with distance from the Galactic center in the
range $25 < r < 80$ kpc: $\sigma(r) = 111 \, (r/40\,
\mathrm{kpc})^{-0.08}\, \kms$.  Spherical Jeans modeling indicates
that the circular velocity profile $V_c(r)$ also falls only slightly
over the same radial range and reaches between 175 and 231 $\kms$ at
80 kpc.  The corresponding enclosed mass $M(80)$ is between $5.7\times
10^{11}\ \Msun$ and $1.0\times 10^{12}\ \Msun$.  A three-component
model for the baryon and dark matter mass distribution gives the total
virial mass of the Galaxy $M_{\rm vir} = (1.6 \pm 0.3)\times 10^{12}\
\Msun$ at the virial radius $R_{\rm vir} = 300$ kpc.

Our inferred mass of the Galaxy is higher than that obtained by
\citet{battaglia_etal05} ($M_{\rm vir} \approx 0.8\times 10^{12}\
\Msun$) and \citet{xue_etal08} ($M_{\rm vir} \approx 1.0\times
10^{12}\ \Msun$) based on the modeling of their radial velocity
datasets.  Our HVS sample contains more objects at $r > 40$ kpc than
the Battaglia et al. and Xue et al. datasets.  Thus we have a stronger
constraint on the shallow slope of the velocity dispersion profile and
we derive a correspondingly larger mass.  Our inferred mass is
consistent with the larger scale measurement by \citet{li_white08}
based on the Andromeda-Milky Way timing argument, $M_{\rm vir} \approx
2.4\times 10^{12}\ \Msun$.  The implied dynamical mass-to-light ratio
of the Galaxy, $M_{\rm vir}/L_V \approx 50$ in solar units, is also
consistent with galaxy-galaxy weak lensing measurements by
\citet{mandelbaum_etal06}, galaxy kinematics modeling by
\citet{more_etal10}, and halo abundance matching modeling by
\citet{moster_etal10}.

Our sample of radial velocities is large enough that the biggest
uncertainty in the mass estimate is not statistical but systematic.
Within the framework of spherical Jeans modeling, the uncertainty is
dominated by the density slope and anisotropy of the tracer
population.  These parameters could be better constrained by future
all-sky surveys of halo BHB stars.  Deeper surveys that target more
distant stars at $r \gtrsim 100$ kpc would be similarly dominated by
uncertainty over the underlying distribution of the tracers.

The validity of spherical Jeans modeling is also limited by the
presence of structure in the distribution of halo stars.  Galactic
stellar halo contains remnants of disrupted satellite galaxies, some
of which are still detectable as tidal streams.  Stars at $\sim 100$
kpc from the Galactic center may not have had enough dynamical times
to reach dynamical equilibrium, further limiting the application of
equilibrium modeling.  A first step in the direction of circumventing
these systematics was taken by \citet{xue_etal08}, who modeled the
motion of particles in realistic simulated halos.  Extension of such
analysis to many different halo realizations using large samples of
observed velocities may reduce the uncertainty over the global mass
distribution in the Galaxy.

\acknowledgements 

We would like to thank Hans-Walter Rix for clarifying discussions and
the organizers of the Sixth Harvard-Smithsonian Sackler Conference,
where this work was completed.  OG is supported in part by NSF grant
AST-0708087.

\bibliography{gc}

\end{document}